\documentclass[conference]{IEEEtran}
\usepackage{blindtext, graphicx}
\usepackage[margin=1in]{geometry}
\usepackage[cmex10]{amsmath}
\usepackage{draftwatermark}

\ifCLASSINFOpdf
\else
\fi

\SetWatermarkText{PREPRINT}
\SetWatermarkScale{1}
\begin{document}

\title{A movie genre prediction based on Multivariate Bernoulli model and genre correlations}

\author{\IEEEauthorblockN{Eric Makita, Artem Lenskiy}
\IEEEauthorblockA{Korea University of Technology and Education\\
1600, Chungjeol-ro, Byeongcheon-myeon,
Dongnam-gu, Cheonan-si, Chungcheongnam-do 31253,\\ Republic of Korea\\
email: ericmakita@koreatech.ac.kr, lensky@koreatech.ac.kr}
}

\maketitle

\begin{abstract}
Movie ratings play an important role both in determining the likelihood of a potential viewer to watch the movie and in reflecting the current viewer satisfaction with the movie. They are available in several sources like the television guide, best-selling reference books, newspaper columns, and television programs. Furthermore, movie ratings are crucial for recommendation engines that track the behavior of all users and utilize the information to suggest items they might like. Movie ratings in most cases, thus, provide information that might be more important than movie feature-based data. It is intuitively appealing that information about the viewing preferences in movie genres is sufficient for predicting a genre of an unlabeled movie. In order to predict movie genres, we treat ratings as a feature vector, apply the Bernoulli event model to estimate the likelihood of a movie’s given genre, and evaluate the posterior probability of the genre of a given movie using the Bayes rule. The goal of the proposed technique is to efficiently use the movie ratings for the task of predicting movie genres. In our approach we attempted to answer the question: "Given the set of users who watched a movie, is it possible to predict the genre of a movie based on its ratings?" Our simulation results with MovieLens 100k data demonstrated the efficiency and accuracy of our proposed technique, achieving 59\% prediction rate for exact prediction and 69\% when including correlated genres.


\end{abstract}


\IEEEpeerreviewmaketitle

\section{Introduction}
Nowadays, web users are no longer simply considered as consumers of information but also as active sources that generate large volume of data online. Consequently, the amount and the diversity of information on the Internet increase exponentially. The whole body of information that confronts the users online can cost them more time and effort without any guarantee of finding what they are looking for. Aiming to solve these problems, researchers in the academe and/or industries have suggested the use of recommender systems \cite{ref1} that overcome the information overload by facilitating search and access to information by providing users with relevant items in the shortest time possible. In this context, items can be of any kind, namely a movie to watch, a soundtrack to listen to, a webpage to visit, or else. Among the widely proposed recommendation techniques, content-based filtering \cite{ref2} \cite{ref3} and collaborative filtering \cite{ref4} \cite{ref5} have been the most famous one in the literature \cite{ref6}. Content-based filtering is done under the assumption that users’ future preferences are similar to those they liked in the past, while collaborative filtering is done under the assumption that if two users have similar preferences in the past, they will have similar preferences in the future. Between them, collaborative filtering is more widely used and therefore attracts more interest from researchers \cite{ref7}\cite{ref8}. Collaborative filtering techniques are rating-oriented and involve the participation of a high number of users who provide fewer ratings than the items they consume. Taking that into consideration, questions such as how to alleviate the data sparsity while increasing the recommendation accuracy are the main concerns in the related works. \\
\indent Recently, some approaches considering factors outside users’ ratings have been proposed in the literature \cite{ref9}. Since recommender systems can naturally be applied in various fields where items are categorized, their associated datasets provide not only users’ ratings but also items’ categories. Based on this information, many recommender system extensions have been made.\\
\indent In this paper, we propose a movie genre prediction model based on users’ ratings. We apply multi-variate Bernoulli model to estimate likelihoods that are used in Naïve Bayes rule to predict movie genres. We also calculated the genre correlations to check if incorrectly predicted genre is correlated with the correctly predicted one. In general, a recommender predicting an item’s category is important in sense that it can complement the item’s categories assigned by a human expert, therefore increasing the user satisfaction by providing surprising recommendations. \\
\indent The proposed approach has the following combination of contributions in order to expand traditional recommender systems : \\
\indent (1) We propose a new approach that expands the traditional recommender systems by predicting the category of an item under evaluation.\\
\indent (2) The multi-variate Bernoulli event model is used to learn a movie's likelihood of belonging to a particular genre.\\
\indent (3) The Bayesian probabilistic reasoning is applied to predict genres. To the best of our knowledge, an item’s category-predicting recommendation is a new attempt. \\
\indent (4) We provide an experimental study of our technique using the MovieLens dataset. Experimental results showed the correctness of our proposed approach.\\
\indent In the remainder of this paper, we present the details of our model with the following organization. Section 2 presents various item genre-oriented recommendations. In section 3, we describe the data model and the mechanism of our proposed technique. Section 4 contains the performance studies. Finally, in section 5, we summarize our work.

\section{Related work}
A large body of research about genres or categories of recommended items has been done in the past.\\
\indent One study proposed a movie recommender system that enhances the accuracy and overcomes the traditional recommendation by factorizing the user-genre matrix instead of the user-item matrix \cite{ref10}. The factorized user-genre matrix model was used to discover latent factors from genres in order to enrich user profiles. In \cite{ref11} content-based filtering utilizing user category-based filtering was proposed to overcome one of the major issues of recommender systems termed as item cold start. Item cold start refers to new items that have not received enough feedback from users, thus could decrease the accuracy of the recommendation. Another example of category-based recommendation is proposed in \cite{ref12}, where authors presented a framework called SEP for overcoming recommender system problems such as cold start and sparsity. The authors in \cite{ref13} proposed a recommender system approach that uses genre information to address not only the coverage of the recommender system algorithms but also the redundancy. Most of the related works focus on designing a new approach to identify similarities between users, while the prediction of movie genres remains understudied to the best of our knowledge. However, it could play an important role in recommending novel items to a user

\section{Proposed method}
The proposed method applies the well-known Bernoulli model for calculating the conditional probability of movie being of a particular genre. To describe our idea clearly, we initially give some definitions used in this paper: users set: $ U $, movie set: $ M $, genre set: $ G $, rating set : $ R $. \\
\indent A movie $ m \in M $ is characterized by a binary feature vector, which components set to 1 if the corresponding user $ u $ rated the movie $ m $ as $ r $ otherwise zero. That is to say:

\begin{equation}
  v_{m,u}(r) = \left\{ \,
   \begin{IEEEeqnarraybox}[][c]{l?s}
     \IEEEstrut
      1 & if $ u \ rated \, movie \ m \ as \ r $, \\
      0 &    $ otherwise $.
      \IEEEstrut
   \end{IEEEeqnarraybox}
\right.
 \label{eq:1}
\end{equation}
Assuming that the ratings of one user do not depend on ratings of other users, the conditional probability of a movie, $ m $ given genre $ g $, is computed according to multi-variate Bernoulli model as follows:

\begin{equation}
\begin{aligned}
P( m \mid g,r ) = & \ \prod\limits_{u \in U} [v_{m,u}(r)P(u \mid g,r) \\
      & \ + (1-v_{m,u}(r))(1-P(u \mid g,r))]
\label{eq:2}
\end{aligned}
\end{equation}

where $v_{m,u}(r)$ is either $0$ or $1$ indicating whether the user $ u $ rated the movie of genre $ g $ as $ r $ or not. A movie can be seen as a collection of multiple independent Bernoulli experiments, one for each user in the user set $ U $ with the probabilities for each of these rating events defined by each component $P( u \mid g,r )$. The probability $P( u \mid g,r )$ defines the probability of user $ u $ given a rating $ r $ to a movie labeled as $ g $ :

\begin{equation}
P( u \mid g,r ) =\frac{1+ \sum_{m \in M}v_{m,u}P( m \mid g,r )}{|U|+\sum_{m \in M}P( m \mid g,r )}
\label{eq:3}
\end{equation}

\indent The probability $P( u \mid g,r )$ can be thought as a user’s preference model towards movie genres. In other words, knowing the genre and the rating, the eq. \ref{eq:3} describes the probability that user $ u $ match a hypothetical user that would rate a movie of genre $ g $ as $ r $. The probability $P( g \mid m,r )$ is $1$ is $m$ is marked only as genre $ g $ otherwise $0$. If $m$ simultaneously belong to $N$ genres, the probability $1/N$ is assigned. To avoid the zero probability that can occur in case when a user $ u $ did not rate a movie $ m $, we added a 1 to the numerator and the total number of users $\lvert U \rvert$ to the denominator according to the Laplace’s law of succession. \\
\indent The posterior probability of a genre, given a movie and the rating is calculated as follows:

\begin{equation}
P( g \mid m,r ) =\frac{P( m \mid g,r ) \cdot P( g \mid r )}{P(m )}
\label{eq:4}
\end{equation}

\indent Given the posterior probability \ref{eq:4}, for each rating we predict a movie $m$ as genre $g$ according to the highest posterior probability as follows:

\begin{equation}
g_{max} = \arg\max_{g \in G} P( g \mid m,r ) 
\label{eq:5}
\end{equation}

\section{Experimental evaluation}

\subsection{Dataset}

We begin with the description of the MovieLens dataset and parameters used in our experiments. \\
\indent We performed our experiments on the MovieLens 100K dataset \cite{ref14},  which contains 1,682 movies, 943 users, and 100,000 ratings that range from range of 1 to 5. In this dataset, each user has rated at least 20 movies. 18 movies' genres were selected here and each movie is assigned to at least one genre. We carried out the experiment by dividing the dataset into two, a training set and a test set. During the training phase rows of the genre matrix and columns in the rating matrix that correspond to the testing set were removed. Thus, the users’ preference models were built only using a portion of the available rated items with known genres selected for testing.

\subsection{Evaluation}

In this section, we demonstrate the correctness of our proposed method.\\
\indent Movies were selected randomly in both of our training and testing approaches. To test the prediction power of the proposed model the following portion of training size were chose: : {1\%, 5\%,10\%,...,75\%, 80\%}. As for the testing set we always used 20\% of the whole set. None of items of testing set were included in the training. \\
\indent The first step in our algorithm is to estimate $P( u \mid g,r )$ based on the user ratings, that is described by eq. \ref{eq:3}. Figure 1, depicts the probability of the user $u$ rating a movie of category $g$ as rating $r = 1$ i.e. user preference models. Figure 2, illustrates  the probability $P( m \mid g,r )$ that is obtained based on the Bernoulli event model (\ref{eq:2}) and plays a role of a likelihood in the Bayesian eq. \ref{eq:4}. The probability of a failure or absence of the event $v_{m,u}(r)$ is given by $(1-v_{m,u}(r))(1-P(u \mid g,r))$.
\indent When the prediction is incorrect, the genre correlation matrix (fig. 3) is used to check whether the incorrectly predicted genre is correlated with the true movie’s genre. If it is correlated $(cor > 0.1)$, we accept this prediction as a prediction of a similar genre. \\
\indent To measure the accuracy of the proposed approach on movie genre recommendations, we plot the predictions based on every rating from one to five. For every training size we repeated the process of randomly selecting training samples 20 times.\\

\begin{figure}[!t]
\centering
\includegraphics[trim=3.5cm 8cm 4cm 7.5cm,width=0.4\textwidth]{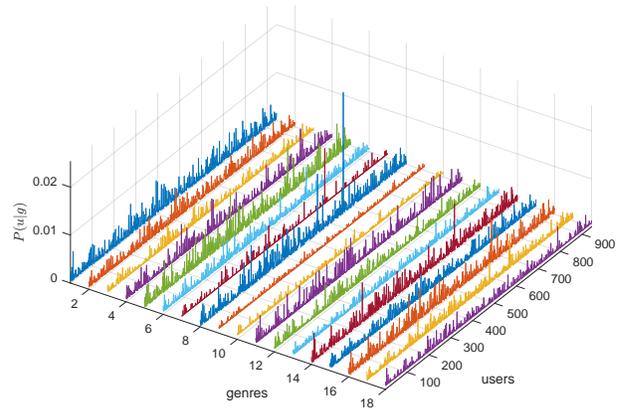}
\caption{Preference models for rating $1$.}
\label{fig1}
\end{figure}

\begin{figure}[!t]
\centering
\includegraphics[trim=3.5cm 8cm 4cm 7.5cm,width=0.4\textwidth]{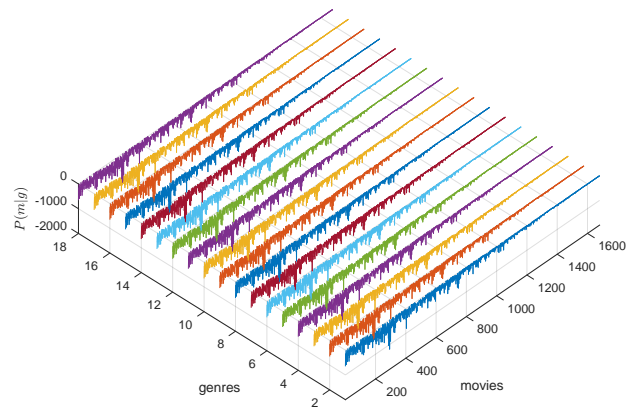}
\caption{Probability for a movie $m$ being rated as $r=1$.}
\label{fig2}
\end{figure}

\begin{figure}[!t]
\centering
\includegraphics[trim=3.5cm 8cm 4cm 7.5cm,width=0.35\textwidth]{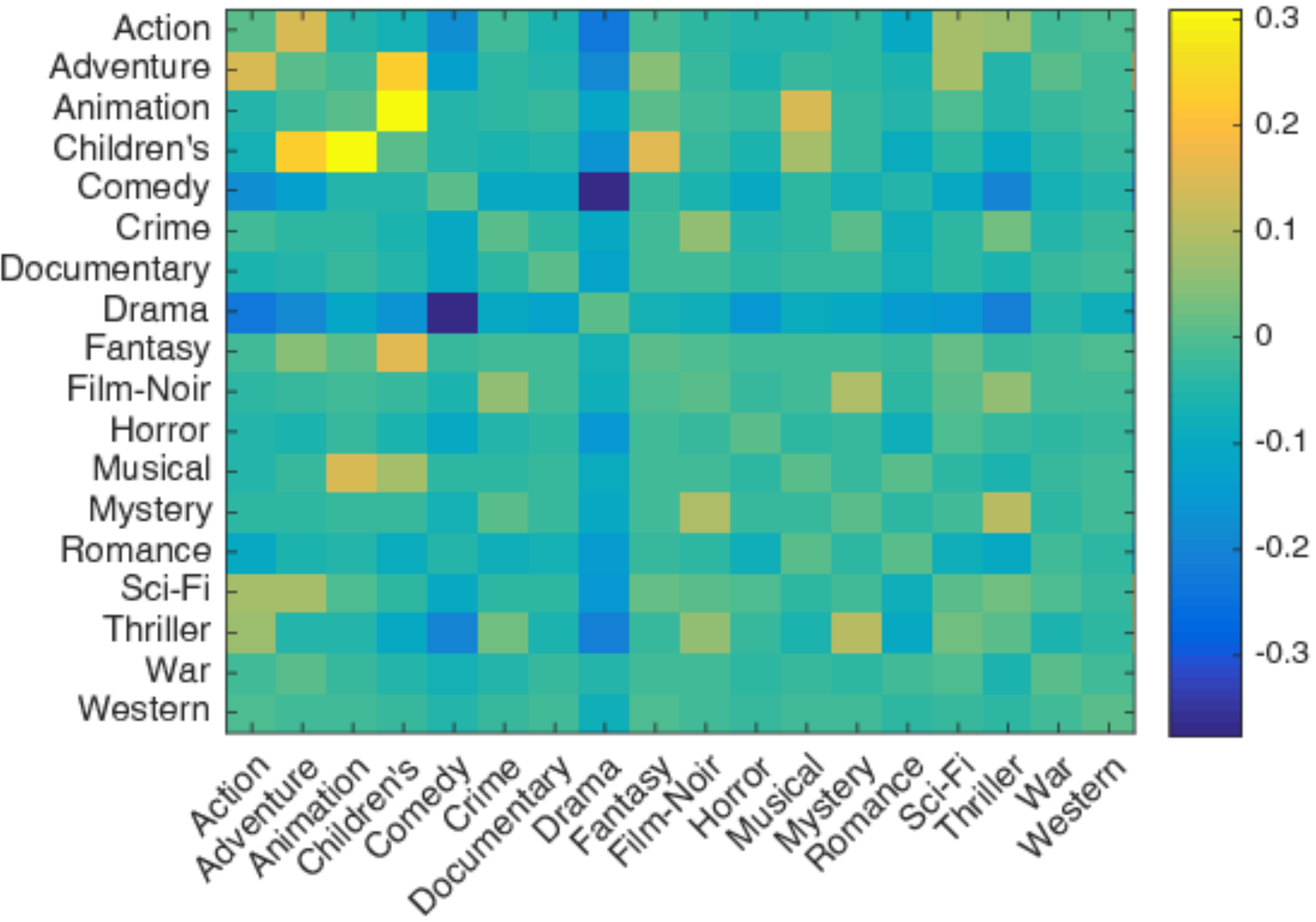}
\caption{Movie correlation matrix.}
\label{fig3}
\end{figure}

\indent Figures 4 to 7 show the prediction rate of our approach with and without including correlated genres.
These plots indicate that our movie category prediction based on the Bayesian model presented in Section 3.0 is effective for predicting genres based on all rating. As the size of the training dataset increases, the prediction accuracy increases as well. Table 1 present the prediction results when 80\% of the whole set was used for the training. \\

\begin{table}[!t]
\centering

\caption{Prediction rate when 80\% of the set used for training}
\label{table1}
\noindent\hrulefill

\resizebox{\columnwidth}{!}{
\begin{tabular}{llllll}
\hline

\begin{tabular}[c]{@{}l@{}} \end{tabular} & 
\begin{tabular}[c]{@{}l@{}}$r = 1$\end{tabular} & 
\begin{tabular}[c]{@{}l@{}}$r = 2$\end{tabular} & 
\begin{tabular}[c]{@{}l@{}}$r = 3$\end{tabular} & 
\begin{tabular}[c]{@{}l@{}}$r = 4$\end{tabular} &
\begin{tabular}[c]{@{}l@{}}$r = 5$\end{tabular}
\\ \hline
exact               & $51.2\pm2.2$ & $51.6\pm2.5$ & $58.8\pm2.6$ & $56.7\pm2.4$ & $52.9\pm2.8$ \\
exact w/\\ similar & $53.6\pm2.2$ & $55.9\pm2.8$ & $69.2\pm2.9$ & $66.6\pm2.4$ & $61.0\pm3.2$ \\

\hline
\end{tabular}
}
\end{table}

\indent It is interesting to note that only 10\% of data used to training is enough to achieve 50\% prediction accuracy. This results shows that Bernoulli multivariate event models can be employed in genre prediction together with Bayesian rule.

\begin{figure}[!t]
\centering
\includegraphics[trim=3.5cm 9cm 5cm 10cm,width=0.4\textwidth]{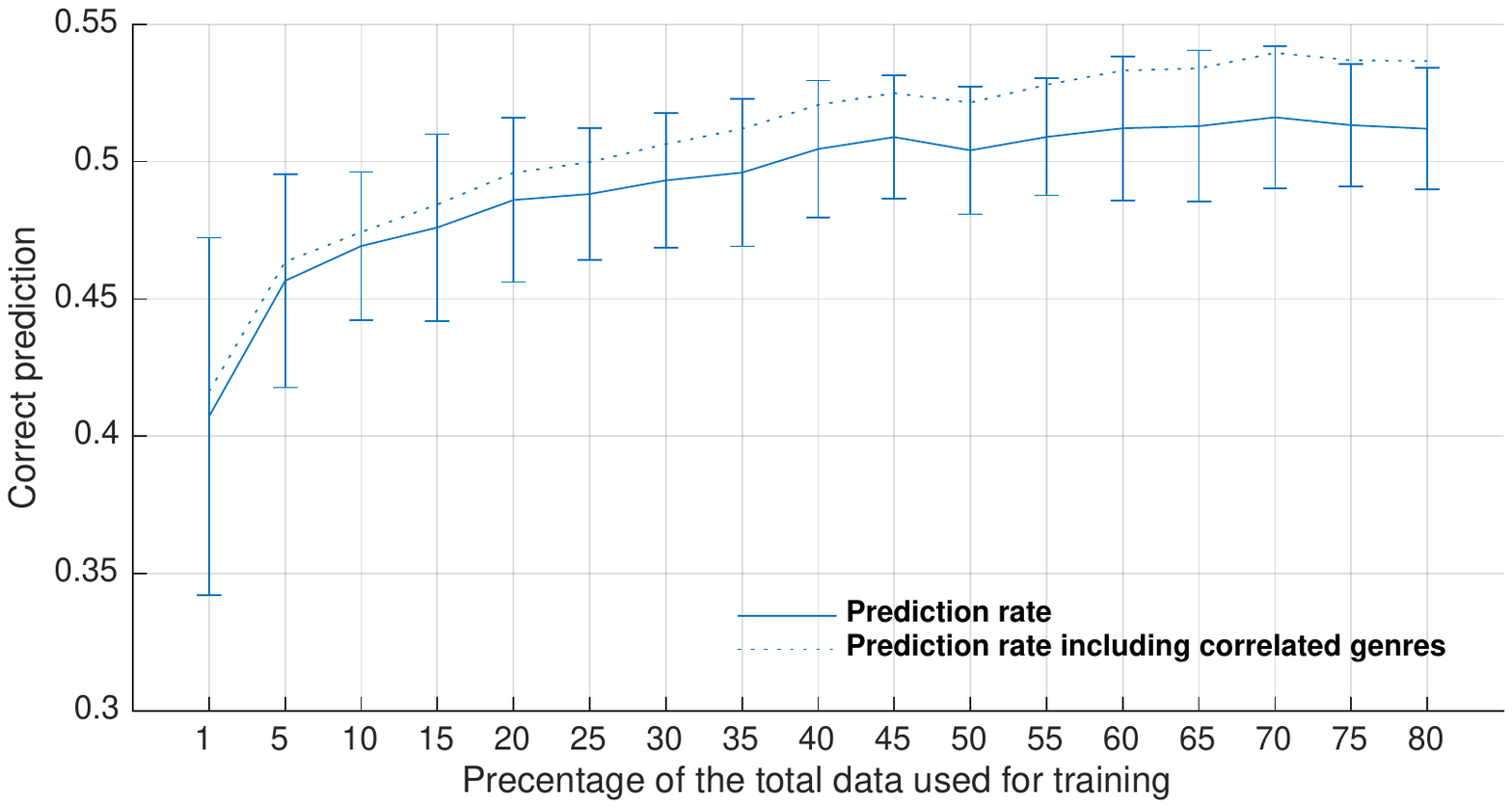}
\caption{Prediction accuracy based on rating $r = 1$.}
\label{fig4}
\end{figure}

\begin{figure}[!t]
\centering
\includegraphics[trim=3.5cm 9cm 5cm 10cm,width=0.4\textwidth]{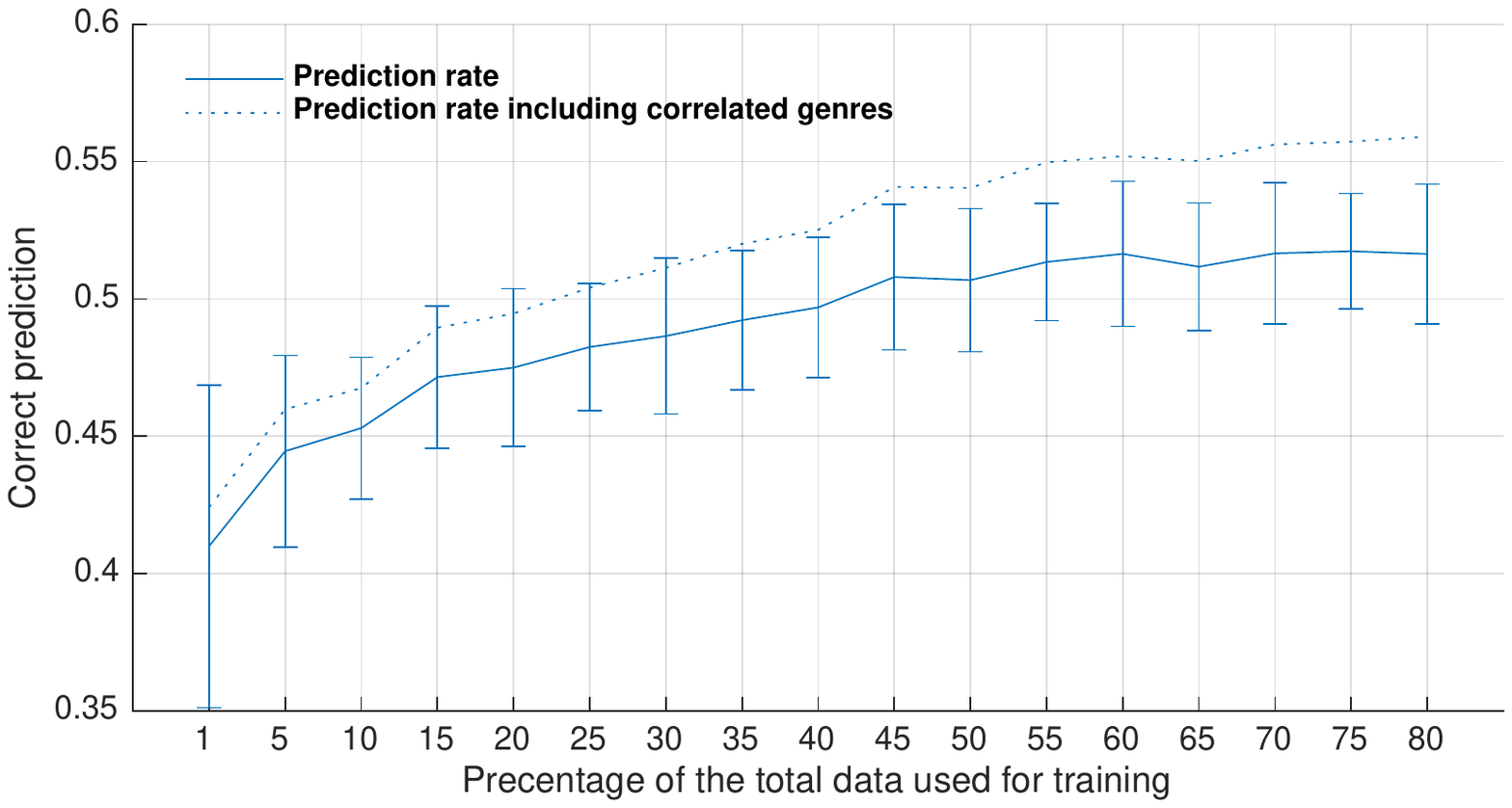}
\caption{Prediction accuracy based on rating $r = 2$.}
\label{fig5}
\end{figure}

\begin{figure}[!t]
\centering
\includegraphics[trim=3.5cm 9cm 5cm 10cm,width=0.4\textwidth]{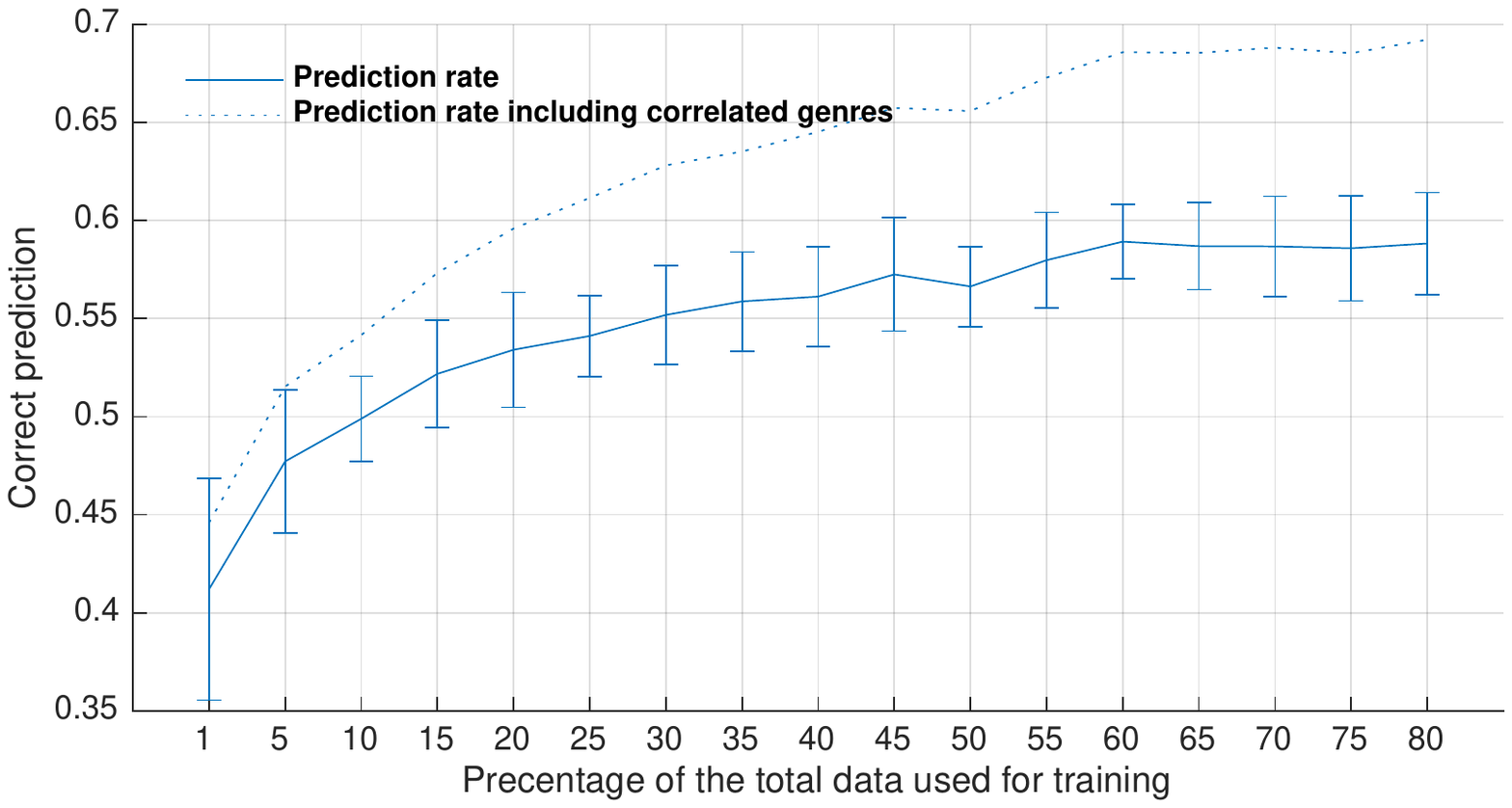}
\caption{Prediction accuracy based on rating $r = 3$.}
\label{fig6}
\end{figure}

\begin{figure}[!t]
\centering
\includegraphics[trim=3.5cm 9cm 5cm 10cm,width=0.4\textwidth]{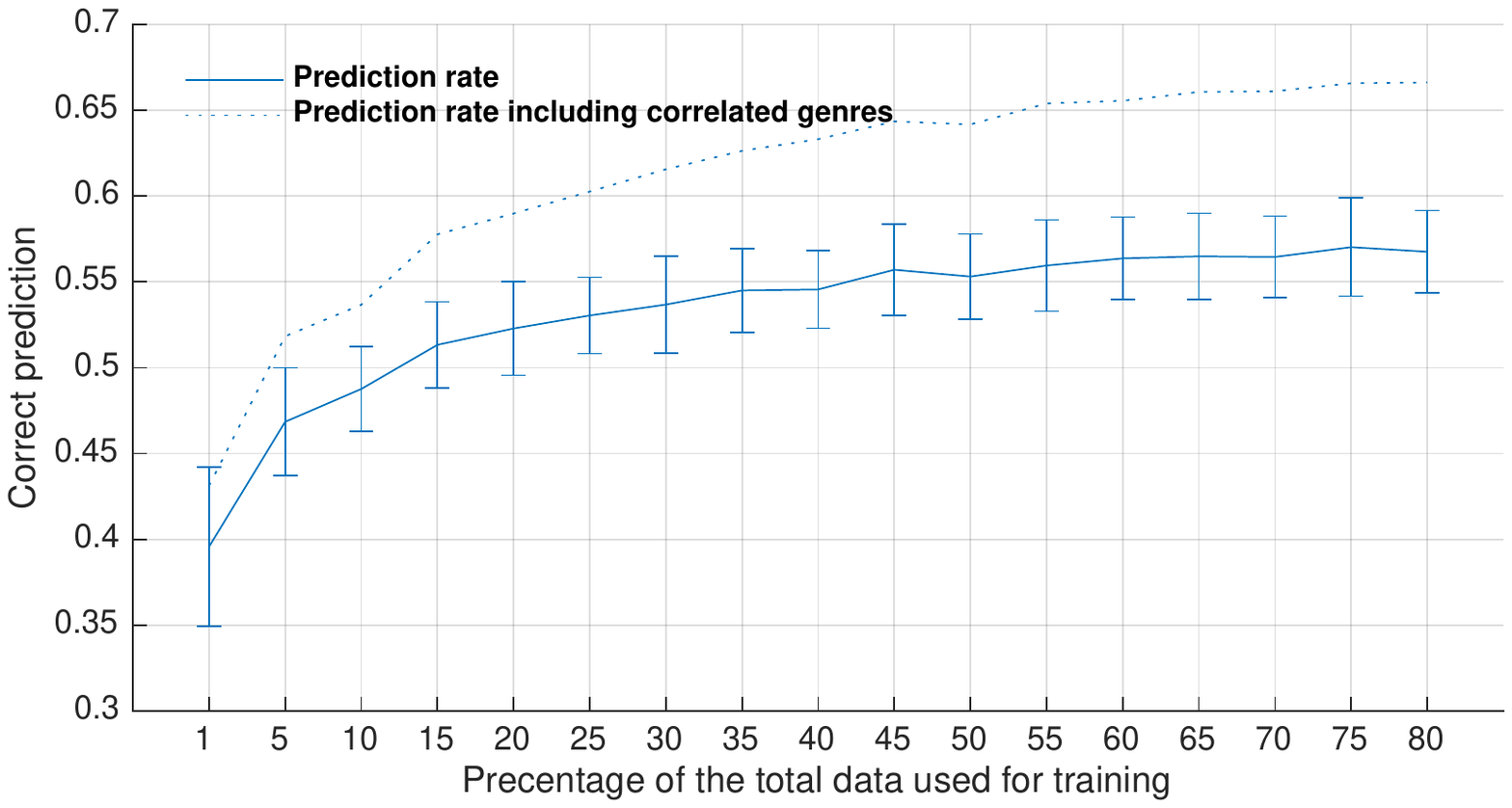}
\caption{Prediction accuracy based on rating $r = 4$.}
\label{fig7}
\end{figure}

\begin{figure}[!t]
\centering
\includegraphics[trim=3.5cm 9cm 5cm 10cm,width=0.4\textwidth]{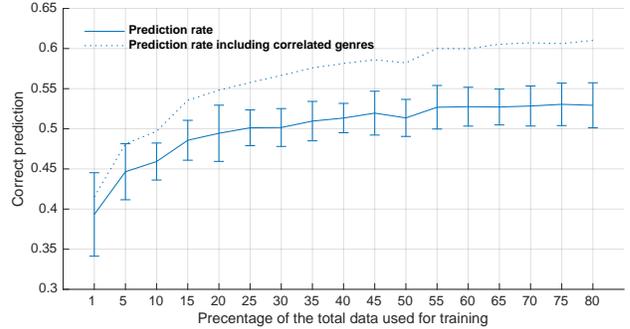}
\caption{Prediction accuracy based on rating $r = 5$.}
\label{fig8}
\end{figure}

\section{Conclusion}
Over the last decade, recommender systems have been successfully applied in various domains such as social networking, online movie viewing website, or e-commerce, etc. Until now, most of the recommender systems that have been proposed and reported in the literature are item rating prediction-oriented. In this paper, we proposed an approach that expands the traditional recommender system algorithms by predicting the category of an item under evaluation rather than predicting its rating. Predicting the category of an item can help increase the accuracy of the recommended items generated by complementing the categories assigned by an expert. \\
\indent To show the correctness of our approach, we conducted an experimental study using MovieLens dataset. The experimental results showed that predicting the category of an item under evaluation can achieve 50\% accuracy rate based on 10\% training set of users’ rating 3.  This finding is deemed valuable in many applications in practice. For instance, it can complement the genres given by experts. It could significantly increase the accuracy and usefulness of recommendations. \\
\indent We also showed in our experimental analysis that predictions based on high ratings do not follow the behavior of predictions based on low ratings. This situation can be seen as an interesting open issue for our future work to tackle, by which we can focus on improving the prediction based on high ratings. Related study may lead to designing new attempts in the field of recommender system.



\ifCLASSOPTIONcaptionsoff
  \newpage
\fi


\end{document}